\documentclass{PoS}

\usepackage{bbm}
\usepackage{graphicx}
\usepackage{multirow}

\newcommand{\et}{{\it et al.}}
\newcommand{\overbar}[1]{\mkern 1.5mu\overline{\mkern-1.5mu#1\mkern-1.5mu}\mkern 1.5mu}

\title{Searching for the $X(3872)$ and $Z_c^+(3900)$ on HISQ lattices}

\ShortTitle{$X(3872)$ and $Z_c^+(3900)$}

\author{Song-Haeng Lee, \speaker{Carleton DeTar}, and Heechang Na \\
        Department of Physics and Astronomy, University of Utah,
        Salt Lake City, Utah, USA\\
        E-mail: \email{song@physics.utah.edu}, \email{detar@physics.utah.edu},
        \email{heena@physics.utah.edu}}

\author{Daniel Mohler\\
       Fermi National Accelerator Laboratory, Batavia, Illinois, USA    \\
       E-mail: \email{mohler@fnal.gov}}

\author{(Fermilab Lattice and MILC Collaborations)}

\abstract{We present preliminary simulation results for the $I = 0$
  charmonium state $X(3872)(1^{++})$ and the $I = 1$ charmonium state
  $Z_c^+(3900)(1^{+-})$.  The study is performed on gauge field
  configurations with 2+1+1 flavors of highly improved staggered sea
  quarks (HISQ) with clover (Fermilab interpretation) charm quarks and
  HISQ light valence quarks.  Since the $X(3872)$ lies very close to
  the open charm $D \bar D^\ast$ threshold, we use a combination of
  $\bar c c$ and $D \bar D^\ast + \bar D D^\ast$ interpolating
  operators.  For the $Z_c^+(3900)$ we use a combination of $J/\psi \pi$
  and $D \bar D^\ast + \bar D D^\ast$ channels.  This is the first
  such study with HISQ sea quarks and light valence quarks.  To this
  end, we describe a variational method for treating staggered quarks
  that incorporates both oscillating and non-oscillating components.
}

\FullConference{The 32nd International Symposium on Lattice Field Theory,\\
		23-28 June, 2014\\
		Columbia University New York, NY}

\begin{document}

\section{Introduction}
In the past decade, many excited charmonium states have been
discovered that cannot be explained within the conventional quark
model. Among these states, the narrow charmonium-like state $X(3872)$
and charged charmonium-like state $Z_c^+(3900)$ have attracted special
attention due to the closeness of the $D \bar D^\ast$ threshold and
their possible four-quark nature.

The $X(3872)$ state with $J^{\rm PC} = 1^{++}$ is one of the better
established mysterious charmonium states found in $B$-meson decays by
both Belle \cite{Choi:2003ue,Adachi:2008te} and CDF
\cite{Acosta:2003zx} and studied with more precison by CDF
\cite{Aaltonen:2009vj}, D0 \cite{Abazov:2004kp}, BABAR
\cite{Aubert:2004ns,Aubert:2008gu}, Belle and LHCb
\cite{Aaij:2011sn,Aaij:2013zoa}.  Its mass is remarkably close to the
$D^0\bar D^{\ast 0}$ threshold -- within $1$~MeV.  The $Z_c^+(3900)$ is a
charged, isospin one charmonium-like structure observed by the BESIII
collaboration \cite{Ablikim:2013mio} as an intermediate resonance in
an analysis of $e^+ e^-$ annihilation into $J/\psi \pi^+\pi^-$ at
$\sqrt{s} = 4260$~MeV. This observation has been confirmed by the
Belle Collaboration \cite{Liu:2013dau} and by Xiao \et\ using data
from the CLEO-c detector \cite{Xiao:2013iha}. However, it has not been
observed in exclusive photoproduction of $J/\psi,\pi$ on protons
\cite{Adolph:2014hba} or in conjunction with $B_0$ decays
\cite{Chilikin:2014bkk,Aaij:2014siy}.  As a charged charmonium-like
structure, it must contain at least four quarks, and tetraquark and
molecular interpretations have been suggested.  See, for example,
\cite{Braaten:2013boa,Braaten:2014qka} and \cite{Wilbring:2013cha}.

Previous lattice studies provide theoretical support for the $X(3872)$
\cite{Prelovsek:2013cra} but not the $Z_c^+(3900)$
\cite{Prelovsek:2013xba,Chen:2014afa,Prelovsek:2014swa}.  Those
studies were carried out on small volumes with unphysically heavy up
and down quarks.  Our ultimate objective is to increase the volume and
work at physical values of all quarks.  To this end the needed gauge
field ensembles with highly improved staggered quarks (HISQ) are
available \cite{Bazavov:2010ru}.  We report here on a preparatory
study, albeit still on a small volume with unphysically heavy up and
down quarks, using the HISQ formulation for the light quarks and
clover (Fermilab interpretation \cite{ElKhadra:1996mp}) for the charm
quark.

\section{Methodology}

We work with the MILC ensemble with lattice spacing approximately
$0.15$ fm and the lattice dimension $16^3 \times 48$, generated in the
presence of highly improved staggered sea quarks (HISQ). The ensemble
contains degenerate up and down sea quarks with masses approximately
1/5 the mass of the strange quark and with strange and charm sea quark
masses at their physical values \cite{Bazavov:2010ru}.

As mentioned above, we use clover charm quarks within the Fermilab
interpretation and HISQ light valence quarks with masses matching the
sea quarks.  To study the $X(3872)$ with $J^{\rm PC} = 1^{++}$, we
choose interpolating operators ${\cal O}_i$ that couple to $\bar c c$
as well as $D \bar D^\ast + \bar D D^\ast$ scattering states.  (We use
abbreviations $cc$ and $D D^\ast$ below.)
\begin{itemize}
  \item $cc$ interpolators ($J^{\rm PC} = 1^{++}$, $I = 0$)
  \begin{displaymath}
      {\bar c} \gamma_5 \gamma_i c , \quad
      {\bar c} \Delta \gamma_5 \gamma_i \Delta c , \quad
      {\bar c} \nabla_k \gamma_5 \gamma_i \nabla_k c , \quad
      {\bar c} \epsilon_{ijk} \gamma_j \nabla_k c , \quad
      {\bar c} \epsilon_{ijk} \gamma_4 \gamma_j \nabla_k c , \quad
      {\bar c} \left| \epsilon_{ijk} \right| \gamma_5 \gamma_j {\cal D}_k c \,.
  \end{displaymath}

  \item $D D^\ast$ interpolators ($J^{\rm PC} = 1^{++}$, $I = 0,1$) 
  \begin{center}
    \begin{tabular}{rcl}
      $(DD)(t,{\bf p}={\bf 0}) $ & : & 
      $[ D^\ast(t,{\bf 0}) {\bar D}(t,{\bf 0}) - {\bar D}^\ast(t,{\bf 0}) D(t,{\bf 0}) ] + f_I \left\{u \leftrightarrow d\right\}$ \\

      $(DD)(t,{\bf p}={\bf 1}) $ & : & 
      $[ D^\ast(t,{\bf -1}) {\bar D}(t,{\bf 1}) - {\bar D}^\ast(t,{\bf 1}) D(t,{\bf -1}) ] $ \\
      & & $ + D^\ast(t,{\bf 1}) {\bar D}(t,{\bf -1}) - {\bar D}^\ast(t,{\bf -1}) D(t,{\bf 1}) ] + f_I \left\{u \leftrightarrow d\right\}$ \\
    \end{tabular}
  \end{center}
\end{itemize}
where $\nabla_k$ is a discrete covariant difference, ${\cal B}_k =
\epsilon_{ijk}\nabla_i\nabla_j$, ${\cal D}_k =
|\epsilon_{ijk}|\nabla_i\nabla_j$, $\Delta = \nabla_k \cdot \nabla_k$,
$f_I = +1$ for $I=0$ and $f_I = -1$ for $I=1$.  On the other hand, for
the $Z_c^+(3900)$ we use the interpolating operators ${\cal O}_i$ that
couple to both $cc\,\pi$ and $DD^\ast$ scattering states with
quantum number $J^{\rm PC} = 1^{+-}$ and $I=1$
\begin{itemize}
  \item $cc$ interpolators ($J^{\rm PC} = 1^{--}$, $I = 0$)
    \begin{displaymath}
      {\bar c} \gamma_i c , \quad
      {\bar c} \gamma_4 \gamma_i \Delta c, \quad
      {\bar c} \nabla_i c , \quad
      {\bar c} \epsilon_{ijk} \gamma_5 \gamma_j \nabla_k c, \quad
      {\bar c} \gamma_5 {\cal B}_k c , \quad
      {\bar c} \gamma_4 \gamma_5 {\cal B}_k c  \,.
    \end{displaymath}
  \item $cc\, \pi$ interpolators ($J^{\rm PC} = 1^{+-}$, $I = 1$)
  \begin{center}
    \begin{tabular}{rcl}
      $ (cc,\pi)(t,{\bf p}={\bf 0})$  & : &  $cc(t,{\bf 0})\pi(t,{\bf 0}) $ \\
      $ (cc,\pi)(t,{\bf p}={\bf 1})$  & : & $cc(t,{\bf -1})\pi(t,{\bf 1}) $ \\
      & & $cc(t,{\bf -1})\pi(t,{\bf 1}) + cc(t,{\bf 1})\pi(t,{\bf -1}) $ \\ 
    \end{tabular}
  \end{center}
  \item $D D^\ast$ interpolators ($J^{\rm PC} = 1^{+-}$, $I = 1$) 
  \begin{center}
    \begin{tabular}{rcl}
      $(DD)(t,{\bf p}={\bf 0}) $ & : & 
      $[ D^\ast(t,{\bf 0}) {\bar D}(t,{\bf 0}) + {\bar D}^\ast(t,{\bf 0}) D(t,{\bf 0}) ] - \left\{u \leftrightarrow d\right\}$ \\

      $(DD)(t,{\bf p}={\bf 1}) $ & : & 
      $[ D^\ast(t,{\bf -1}) {\bar D}(t,{\bf 1}) + {\bar D}^\ast(t,{\bf 1}) D(t,{\bf -1}) ] $ \\
      & & $ + D^\ast(t,{\bf 1}) {\bar D}(t,{\bf -1}) + {\bar D}^\ast(t,{\bf -1}) D(t,{\bf 1}) ] - \left\{u \leftrightarrow d\right\}$ \\
    \end{tabular}
  \end{center}

\end{itemize}
Each charmed meson interpolating operator is given by
\begin{eqnarray}
  D(t,{\bf p}) = \sum_x e^{i{\bf p}\cdot{\bf x}} {\bar u}({\bf x},t) \gamma_5 c({\bf x}, t) & {\rm ,} &
  D^\ast(t,{\bf p}) = \sum_x e^{i{\bf p}\cdot{\bf x}} {\bar u}({\bf x},t) \gamma_i c({\bf x}, t) 
 \end{eqnarray}
and stochastic and smeared-stochastic sources are used throughout. 

\section{Staggered variational method}

To extract the discrete energy spectrum $E_n$ of the various
scattering states, we use a variational approach
\cite{Michael:1985ne,Luscher:1990ck,Blossier:2009kd}. The extension to
staggered quarks is described in \cite{LeeDeTar}.  When the hadronic
correlator involves staggered fermions, the multi-exponential
expansion of the correlator includes terms that oscillate in time:
\begin{equation}
  C_{ij}(t) = \langle {\cal O}_i(0) {\cal O}_j(t) \rangle
            = \sum_n s_n(t)Z_{in} Z_{jn}^\ast \frac{\exp(-E_n t)}{2E_n} \,,
\end{equation}
where $s_n(t) = 1$ or $-(-)^t$ for nonoscillating and oscillating states.  
In terms of a pseudo-transfer matrix $T$ with eigenvalues $\pm \exp(-E_n)$
\begin{equation}
  C(t) = Z T^t g(2M)^{-1} Z^\dag \,,
\end{equation}
where $g$ is diagonal with $g_{nn} = 1$ for nonoscillating and $-1$ for
oscillating states, and $M$ is a diagonal matrix with $M_{nn} = E_n$.
We obtain the generalized eigenvalue problem:
\begin{equation}
  C(t) V = T^{t-t_0} C(t_0) V \,,
  \label{eq:StagVar:GEVP}
\end{equation}
where the eigenvector $V = Z^{\dag -1}$. With a sufficiently complete 
interpolating operator basis and a high reference time $t_0$, we get the  
eigenvalues, 
\begin{equation}
  \lambda_n(t,t_0) = s_n(t) \exp[-E_n(t-t_0)] \,.
  \label{eq:StagVar:Eigenvals}
\end{equation}
However, in practice, if the basis is not sufficiently complete and
$t_0$ is not sufficiently high, $\lambda_n(t,t_0)$ receives
contribution from higher states and often from opposite parity states,
so we fit to
\begin{eqnarray}
  \lambda_n(t,t_0) &\approx& [1-a_n(t_0)] s_n(t-t_0) e^{-E_n (t - t_0)} + b_n(t_0) s_n(t-t_0)e^{-\bar E_n (t - t_0)} + \nonumber \\
  &+& c_n(t_0) s_n^\prime(t-t_0) e^{-E_n^\prime (t - t_0)} 
  + d_n(t_0) s_n^\prime(t-t_0) e^{-\bar E_n^\prime (t - t_0)} \, .
  \label{eq:StagVar:fiteigenval}
\end{eqnarray}
where $s_n^\prime$ oscillates if $s_n$ does not, or vice versa.

\section{Results}

\subsection{$X(3872)$}
\begin{figure}
  \begin{center}
  \includegraphics[width=0.65\textwidth]{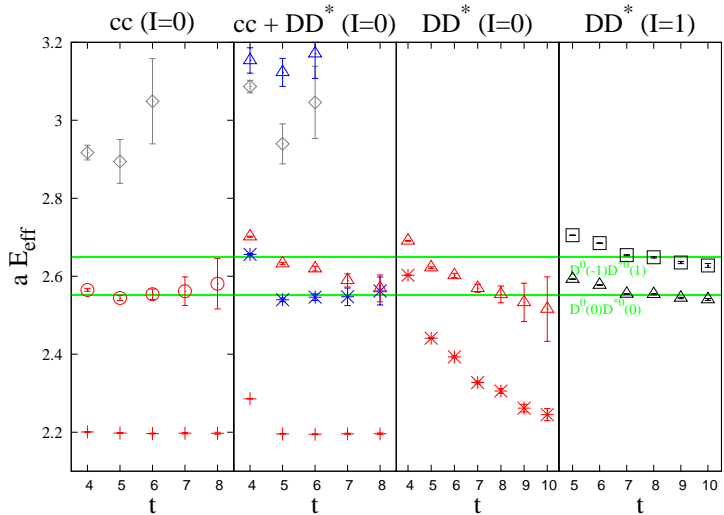}
  \caption{Effective masses in lattice units from the lowest few
    eigenvalues in the $X(3872)$ study.  Each panel shows the result
    of including a different set of interpolating operators. The green
    lines correspond to the energies of non-interacting ${\bar D}({\bf
      p}) D(-{\bf p})$ scattering states. The lower one represents
    ${\bar D}({\bf 0}) D(-{\bf 0})$ and upper, ${\bar D}({\bf 1})
    D(-{\bf 1})$.  The symbols represent effective masses for
    different sets of interpolating operators, panel (a): $cc$ set only,
    (b): combining $cc$ and $D D^\ast$, (c): $D D^\ast$ with
    isospin 0, and (d): $D D^\ast$ with isospin $1$.}

  \label{fig:X3872effmass}
  \end{center}
\end{figure}
\begin{figure}
  \begin{center}
  \includegraphics[width=0.55\textwidth]{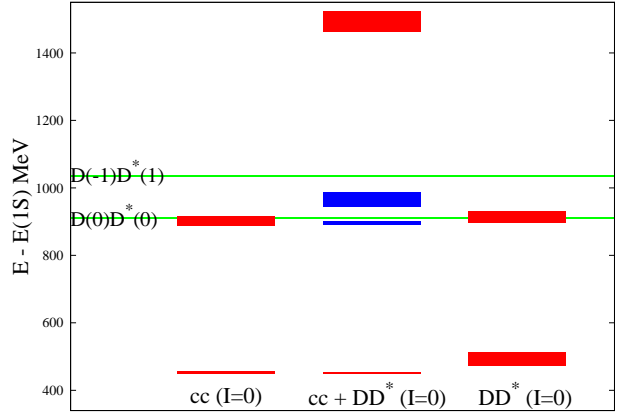}
  \caption{Energy splittings between $E_n$ and $\overbar{1S} =
    \frac{1}{4}(M_{\eta_c} + 3 M_{J/\psi})$, the spin-averaged $1S$
    charmonium masses.  The towers of states are from the same
    operator bases as the first three panels in
    Fig.~\protect\ref{fig:X3872effmass}.  Left: the separate
    $\chi_{c1}(1P)$ and $\chi_{c1}(2P)$ states from $cc$ operators.
    Middle: combined $cc$ and $DD^\ast$ operators.  Right: states
    from the $DD^\ast$ $I = 0$ operators.  The lower blue bar represents
    the $X(3872)$ candidate.}
  \label{fig:X3872spectrum}
  \end{center}
\end{figure}
The resultant effective masses and spectrum in this preliminary study
are shown in Fig.~\ref{fig:X3872effmass} and
Fig.~\ref{fig:X3872spectrum}, respectively.  In the isotriplet channel
we do not find a candidate for the $X(3872)$.  The levels observed are
apparently only discrete scattering state of $DD^\ast$ which
inevitably appear on the lattice.  Any isotriplet character for the
$X(3872)$ would presumably arise after breaking the degeneracy of the
up and down quarks.

The isosinglet channel includes mixing with the $\bar c c$ states.
Our choices of quark masses resulted in a degeneracy
between the unmixed $\chi_{c1}(2P)$ and the unmixed $D D^\ast$
threshold.  With all interpolating operators included, level repulsion
results in the weakly bound state represented by the lower blue bar,
our candidate $X(3872)$.  The upper blue bar can be interpreted as a
scattering state shifted up due to the large negative scattering
length. This shallow bound state scenario on the lattice has been
confirmed in deuteron studies \cite{Yamazaki:2011nd,Beane:2013br}.
Our results agree qualitatively with those of the pioneering lattice
studies of the $X(3872)$ by Prelovsek and Leskovec
\cite{Prelovsek:2013cra} using clover valence and sea quarks
throughout.
\begin{table}
  \caption{ Energy levels for the $cc+DD^\ast$ operator set.  The level $e_1$
    (lower blue bar in Fig.~\protect\ref{fig:X3872spectrum})
    corresponds to the $X(3872)$ candidate with a splitting of
    $13(6)$ MeV relative to the $DD^\ast$ threshold with our
    unphysical lattice parameters.}
  \label{tab:X3872spectrum}
  \begin{center}
  \begin{tabular}{|c|c|c|}
    \hline 
    & & $E_n - \overbar{1S}$ (MeV) \\
    \hline
    \multirow{2}{*}{Non-interacting} & ${\bar D}({\bf 0}) D({\bf 0})$ & 910(2) \\
    & ${\bar D}({\bf 1}) D(-{\bf 1})$ & 1036(3) \\
    \hline
    \multirow{4}{*}{Interacting} &  $e_0$  &  $452(2)$ \\
    &  $e_1$  &  $897(6)$ \\
    &  $e_2$  &  $966(21)$ \\
    &  $e_3$  &  $1494(30)$ \\
    \hline
  \end{tabular}
  \end{center}
\end{table}

\subsection{$Z_c^+(3900)$}

\begin{figure}
  \centering
  \includegraphics[width=0.55\textwidth]{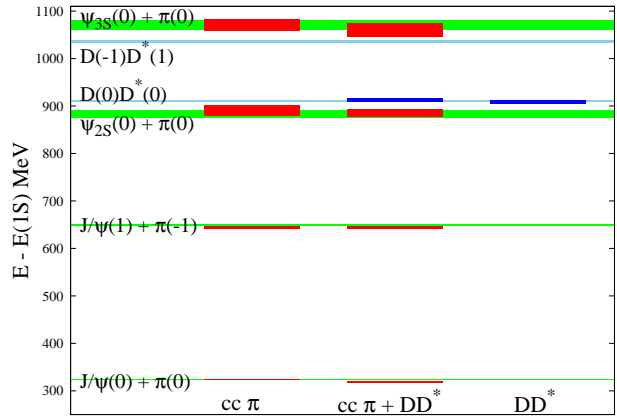}
  \caption{Same as Fig.~\protect\ref{fig:X3872spectrum}, but for three
    choices for the $1^{+-}$ operator basis proposed for the
    $Z_c^+(3900)$.  Left tower: $cc\,\pi$ operators only, middle
    tower: combined operators $cc\,\pi + D D^\ast$ and right tower: $D
    D^\ast$ operators only.  The horizontal green lines represent
    energy levels of the non-interacting $cc\,\pi$
    states and the horizontal blue lines, $D D^\ast$.}
  \label{fig:Z3900spectrum}
\end{figure}

Figure~\ref{fig:Z3900spectrum} shows the energy splittings in the
various $1^{+-}$ channels.  The mixing is evidently too weak to produce a
state distinct from the noninteracting scattering states, in agreement
with \cite{Prelovsek:2013xba,Prelovsek:2014swa}.

\section{Conclusions and Outlook}

In this exploratory study, we find a candidate $X(3872)$ state with an
energy level $13(6)$ MeV below the $DD^\ast$ threshold in the $cc +
DD^\ast$, $I=0$ operator set.  Since the rms separation of the $D$ and
$D^*$ mesons could be quite large ($\sim 6$ fm) \cite{Braaten:2009zz},
we intend to repeat the calculation on a larger lattice with physical
light quark masses. We were unable to observe a candidate
$Z_c^+(3900)$ state, although future calculations with a larger
interpolating operator basis may be able to resolve this state.

C.D., S.-H.L., and H.N. are supported by the U.S. National Science
Foundation under grant NSF PHY10-034278 and the U.S. Department of
Energy under grant DE-FC02-12ER41879.  Fermilab is operated by
Fermi Research Alliance, LLC, under Contract No.~DE-AC02-07CH11359
with the United States Department of Energy.  Calculations were done
on the FNAL cluster, Utah CHPC cluster, and NERSC.

\providecommand{\href}[2]{#2}

\end{document}